\documentclass[aip,preprint]{revtex4-1}

\usepackage{graphicx}
\usepackage{dcolumn}
\usepackage{bm}
\usepackage{amsmath,amssymb}

\def\lsim{\mathrel{\rlap{\lower4pt\hbox{\hskip1pt$\sim$}}
    \raise1pt\hbox{$<$}}}                
\def\gsim{\mathrel{\rlap{\lower4pt\hbox{\hskip1pt$\sim$}}
    \raise1pt\hbox{$>$}}}                

\begin{document}


\title{Dimensionality and heat transport in Si-Ge superlattices}
\author{Ivana~Savi\'c}
\email{ivana.savic@tyndall.ie}
\affiliation{
Department of Chemistry, University of California Davis, Davis, California 95616, USA
}
\author{Davide Donadio}
\affiliation{
Max Planck Institute for Polymer Research, Mainz, Germany
}
\author{Fran{\c c}ois Gygi}
\affiliation{
Department of Computer Science, University of California Davis, Davis, California 95616, USA
}
\author{Giulia Galli}
\affiliation{
Department of Chemistry and Department of Physics, University of California Davis, Davis, California 95616, USA
}


\begin{abstract}

 We investigated how dimensionality affects heat transport in Si-Ge
superlattices by computing the thermal conductivity of planar superlattices
and arrays of Ge nanowires and nanodots embedded in Si. We studied
superlattices with $\sim$10 nm periods using a fully atomistic Monte Carlo solution 
of the Boltzmann transport equation in the relaxation time
approximation. We found that for periods larger than 4 nm,
the room temperature cross-plane conductivity of planar superlattices with
equally thick Si and Ge layers is larger than that of their nanowire
and dot counterparts of similar sizes (up to 100\%), while the trend is 
reversed below 4 nm.

\end{abstract}

\maketitle

In recent years, nanostructuring has emerged as a promising strategy to tailor the properties of semiconducting materials for energy applications~\cite{nmat4-366}. In the field of thermoelectricity much effort has been dedicated to manipulate, at the nanoscale, the thermal transport properties of simple semiconductors, in particular Si, Ge and their composites~\cite{advmat19-1043}. It is now well established that SiGe superlattices~\cite{apl70-2957,slsmicro28-199,apl80-1737,prb67-165333,jes152-G432,apl92-053112,apl93-013112} and nanocomposites~\cite{nl12-4670,apl93-193121} may exhibit a lower thermal conductivity ($\kappa$) than bulk SiGe. In addition it was found that dimensionality may affect the value of $\kappa$, e.g. Pernot et al.~\cite{nmat9-491} reported a ratio as high as three between the conductivity of Si-Ge nanodot (ND) multilayers and that of planar superlattices. However the effect of dimensionality on heat transport in SiGe, and in general in semiconductor nanomaterials, is not fully understood. Experimentally it is difficult to separate the effect of dimensionality from that of interface roughness and defects, and it has long been prohibitive from a computational standpoint, to carry out atomistic calculations for sufficiently large systems, so as to systematically investigate dimensionality effects. 

Studies of how dimensionality affects the thermal conductivity of superlattices (SLs) have so far been conducted using non-atomistic models on samples with completely diffuse (rough) interfaces~\cite{advmat19-1043,prb69-195316,jht130-042410,jjap50-035201}, where lifetimes and group velocities were approximated with those of the bulk and averaged over the entire frequency range. These models predicted that the $\kappa$ of SLs is uniquely determined by the interface density (interfacial area per unit volume) regardless of their dimensionality. Although such predictions may be valid in the macroscopic limit, they are not expected to hold for nanostructures with characteristic sizes of $\sim 10$~nm, where phonon dispersions
and lifetimes become markedly different from those of the bulk. 
So far, atomistic studies of SLs have mostly focused on two-dimensional (planar) SLs~\cite{prb70-081310,nl12-5135}. Only a few molecular dynamics investigations on nanowire (NW) composites have been reported~\cite{jpd43-135401,jpd42-095416,apl100-091903}.

In this Letter,
we investigated the influence of dimensionality on the thermal conductivity of Si-Ge superlattices with period lengths up to $70$ nm. We developed a fully atomistic Monte Carlo (MC) method to solve the Boltzmann transport equation (BTE) in the relaxation time approximation (RTA)~\cite{srivastava}, which allows one to treat systems ten times larger than using exact integration techniques, and with a reciprocal space resolution
improved by an order of magnitude.  We considered planar SLs with Si and Ge layers of equal thickness (two-dimensional SLs, 2D), and periodically repeated Ge NWs (1D) and NDs (0D) embedded in Si of similar characteristic size, grown in the [001] direction.
We found that at 300 K reducing the dimensionality of SLs decreases the cross-plane thermal conductivity $\kappa_{\perp}$ for periods $L$ larger than $4$ nm, while the trend is reversed for shorter periods. 
We showed that this cross-over is not simply related to changes in the interface density; 
 it results from a delicate interplay
 of changes in the phonon density of states, group velocities and lifetimes, which are not correctly described using models based on 
bulk data~\cite{advmat19-1043,prb69-195316,jht130-042410,jjap50-035201}. These findings emphasize the importance of accurate, microscopic descriptions when predicting the properties of nanomaterials.

Within the BTE-RTA approach, the thermal conductivity is given by $\kappa=\sum_{{\bf q},s}c_{{\bf q},s}v_{{\bf q},s}^2\tau_{{\bf q},s}/NV$~\cite{srivastava,prb79-064301}, where ({\bf q},s) denotes a phonon mode with reciprocal space vector {\bf q} and branch index $s$, $\omega_{{\bf q},s}$ is its frequency, $c_{{\bf q},s}$ the heat capacity, $v_{{\bf q},s}=\text{d}\omega_{{\bf q},s}/\text{d}{\bf q}$ the group velocity, and $\tau_{{\bf q},s}$ the lifetime. $V$ is the supercell volume and $N$ is the ${\bf q}$ point grid size. We calculated phonon lifetimes taking into account the contribution of 
three-phonon processes~\cite{srivastava,prb79-064301}:
\begin{eqnarray}
\tau_{{\bf q},s}^{-1}=\frac{\pi\hbar}{4N}\sum_{{\bf q'},s'}\sum_{{\bf q''},s''}\delta_{{\bf G},{\bf q+q'+q''}}\frac{|V_3({\bf q}s,{\bf q'}s',{\bf q''}s'')|^2}{\omega_{{\bf q},s}\omega_{{\bf q'},s'}\omega_{{\bf q''},s''}}\nonumber\\
\left[0.5(1+n_{{\bf q'},s'}+n_{{\bf q''},s''})\delta(\omega_{{\bf q},s}-\omega_{{\bf q'},s'}-\omega_{{\bf q''},s''})\right.\nonumber\\\left.
+(n_{{\bf q'},s'}-n_{{\bf q''},s''})\delta(\omega_{{\bf q},s}+\omega_{{\bf q'},s'}-\omega_{{\bf q''},s''})\right],
\end{eqnarray}
where $\hbar$ is Planck's constant, {\bf G} a reciprocal lattice vector, $V_3({\bf q}s,{\bf q'}s',{\bf q''}s'')$ are the three-phonon coupling elements~\cite{srivastava,prb79-064301}, and $n_{{\bf q},s}$ is the mode occupation. 

We used MC integration techniques~\cite{numericalrecipes} to calculate both the thermal conductivity and the phonon lifetimes. We randomly sampled the phonon modes ({\bf q},s) to compute $\kappa$. For each of these modes, we calculated the lifetime by randomly choosing the modes ({\bf q'},s') and ({\bf q''},s'') that interact with ({\bf q},s)  (see Eq. (1)). In both cases, we selected as many points as necessary to obtain desired statistical error bars ($\sim 3\--10$\%).

We used importance sampling~\cite{numericalrecipes} to reduce the variance and accelerate the MC integration.
In particular, to construct the distribution function that samples the modes ({\bf q},s) in the $\kappa$ calculation,
we exploited the fact that  for the systems considered here $\tau_{{\bf q},s}\sim \omega_{{\bf q},s}^b$, with $b\sim -1.4$
for frequencies $\leq 200$ cm$^{-1}$ (i.e. those providing the largest contribution to $\kappa$). Even if we use a much less accurate estimate of the $\tau(\omega)$ dependence (e.g. $b=-0.5$ or $-3$ in $\tau\sim \omega^b$), the accuracy of the integration technique is not affected, although the efficiency decreases~\cite{footnote_si}. In the computation of phonon lifetimes with importance sampling, we used the normalized two phonon density of states as the distribution function selecting ({\bf q'},s') and ({\bf q''},s''). Delta functions in the two phonon density of states and in Eq.~(1) were calculated with the linear tetrahedron method~\cite{pssb54-469}.
We used the Tersoff interatomic potential for Si and Ge~\cite{prb39-5566}, 
nevertheless the MC technique developed here can be straightforwardly implemented for first principle Hamiltonians. 

The described technique represents the first fully atomistic MC method to compute the thermal conductivity within the BTE-RTA framework, where phonon dispersions and three-phonon lifetimes of nanoscale materials are calculated fully atomistically, without introducing any approximations or fitting parameters.
The use of the double MC integration outlined above allowed us to calculate $\kappa$ for systems that are one order of magnitude larger than those accessible using exact integration.
For example, we computed $\kappa$ of planar SLs with 1024 atoms, and NW and ND SLs with  2048 and 4096 atoms, respectively  (see Fig.~3), while the exact integration was limited 
(cpu wise) to systems with 128 (SL), 288 (NW) and 512 (ND) atoms (see Fig.~2(a)). Furthermore, using the MC method, we could use reciprocal space grids that are 8 times larger than those employed in the exact integration, thus ensuring the convergence of our results~\cite{footnote_si}. 
\begin{figure}
\includegraphics[width=0.6\textwidth]{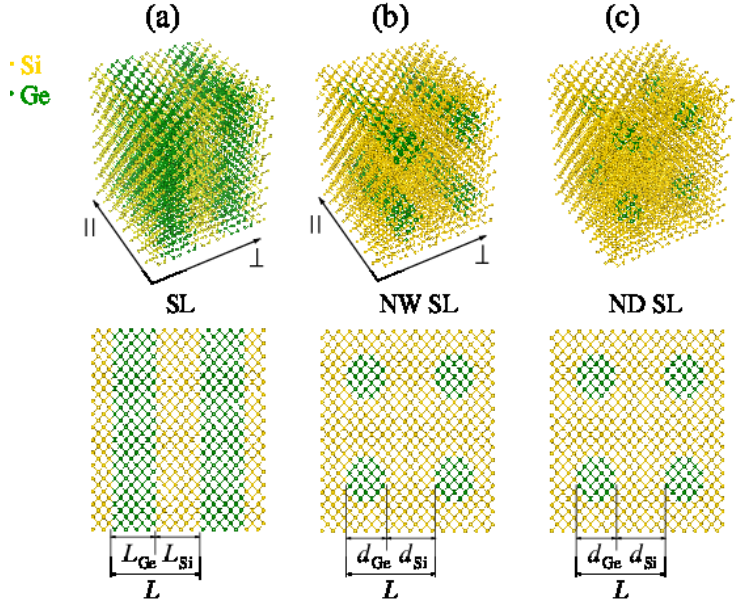}
\caption{Atomic structure of Si-Ge superlattices investigated in this work: (a) planar superlattice (SL), (b) nanowire superlattice (NW SL), and (c) nanodot superlattice (ND SL). 
 The heat propagation directions parallel ($\parallel$, in-plane) and perpendicular ($\perp$, cross-plane) to the Ge layers and NWs are also shown (they are equivalent for ND SLs).}
\label{fig1}
\end{figure}

We focused on SLs grown in the [001] direction (see Fig.~1):
(a) planar SLs with Si and Ge layers of equal thickness ($L_{\text{Si}}=L_{\text{Ge}}$), (b) NW SLs with cylindrical NWs, and (c) ND SLs with spherical NDs.
The SL period is defined as $L=L_{\text{Si}}+L_{\text{Ge}}$ (a), and $L=d_{\text{Si}}+d_{\text{Ge}}$ (b,c), where $d_{\text{Si}}=d_{\text{Ge}}+2$ atomic layers; $d_{\text{Si}}$ is the distance between Ge NWs and NDs in (b) and (c), respectively, and $d_{\text{Ge}}$ is their diameter.
We calculated
the cross-plane and the in-plane conductivity ($\kappa_{\parallel}$) for planar and NW SLs (corresponding to heat propagating in the direction perpendicular and parallel to the Ge layers and NWs, respectively; these directions are equivalent for ND SLs).

We first verified that for small systems the MC method reproduces the same results as obtained using exact integration. Excellent agreement between the two methods is demonstrated in Fig.~2(a) showing the thermal conductivity at 300 K as a function of the period $L$ for planar (upper panel), NW (middle panel) and ND SLs (lower panel). Lifetimes as a function of frequency for $L=22$~\AA~and the same temperature are shown in Fig.~2(b);
those obtained with MC and exact calculations are basically identical. 
Fig.~2(b) also illustrates that low frequency phonons, which largely determine the value of $\kappa$, are much better sampled using MC with importance sampling than high frequency ones.
\begin{figure}
\includegraphics[width=0.65\textwidth]{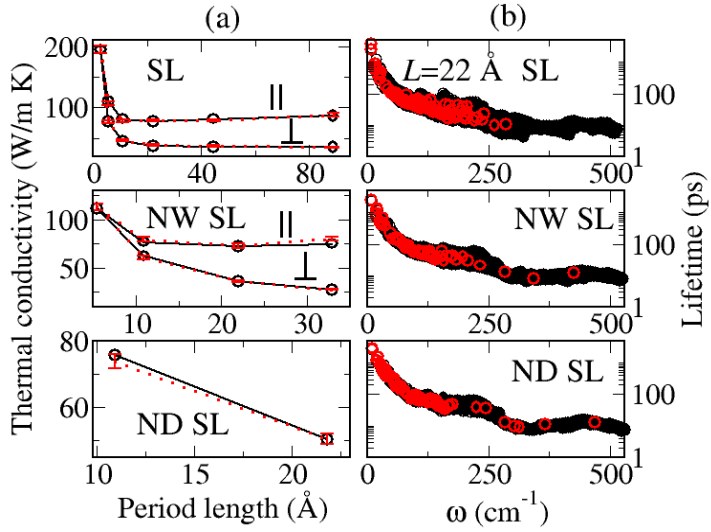}
\caption{(a) The cross- ($\parallel$) and in-plane ($\perp$) thermal conductivity at 300 K as a function of the superlattice period $L$. (b) The lifetimes dependence on the phonon frequency for $L=22$ \AA~and 300 K. Upper, middle and lower panels correspond to planar (SL), nanowire (NW SL) and nanodot (ND SL) superlattices, respectively. The black solid line and circles represent the exact Boltzmann transport solution in the relaxation time approximation, while the red dashed line and circles show the Monte Carlo results.}
\label{fig2}
\end{figure}

The $\kappa$ of Si-Ge SLs as a function of 
dimensionality, computed using the MC method, is reported in Fig.~3. The figure shows the ratios $\kappa_{\perp}/\kappa_{\text{Si}}$ and $\kappa_{\parallel}/\kappa_{\text{Si}}$ at $300$ K (where $\kappa_{\text{Si}}=310.1$ W/m K is the conductivity of bulk Si~\cite{pccp}) for planar (black line), nanowire (red line) and nanodot SLs (green line), as a function of $L$. (Also, $\kappa_{\text{Ge}}=84.6$ W/m K is our calculated $\kappa$ of bulk Ge at $300$ K~\cite{pccp}). 
Both $\kappa_{\perp}$ and $\kappa_{\parallel}$ of planar SLs (black solid and dashed lines in Fig.~3) exhibit a minimum as a function of $L$, consistent with experiment~\cite{slsmicro28-199} and theory~\cite{prl84-927}.
Our MC integration scheme allowed us to probe, at the atomistic level,
larger periods for planar SLs than previously investigated~\cite{prb70-081310,nl12-5135}, where we could detect that $\kappa$ increases with $L$. For sufficiently large periods, SLs should behave as a series of resistors that correspond to bulk Si and Ge and their boundaries~\cite{prl84-927}. Consequently, $\kappa$ is expected to increase with $L$, and unlike previous 
atomistic 
studies on planar SLs~\cite{prb70-081310,nl12-5135}, our BTE-MC calculations capture the onset of this 
behavior.
In the case of NW SLs, we found a minimum in $\kappa_{\parallel}$, similar to planar SLs (see Fig.~3). On the other hand, for ND and NW SLs, $\kappa_{\perp}$ decreases as a function of $L$, for all investigated $L$. 
\begin{figure}
\includegraphics[width=0.6\textwidth]{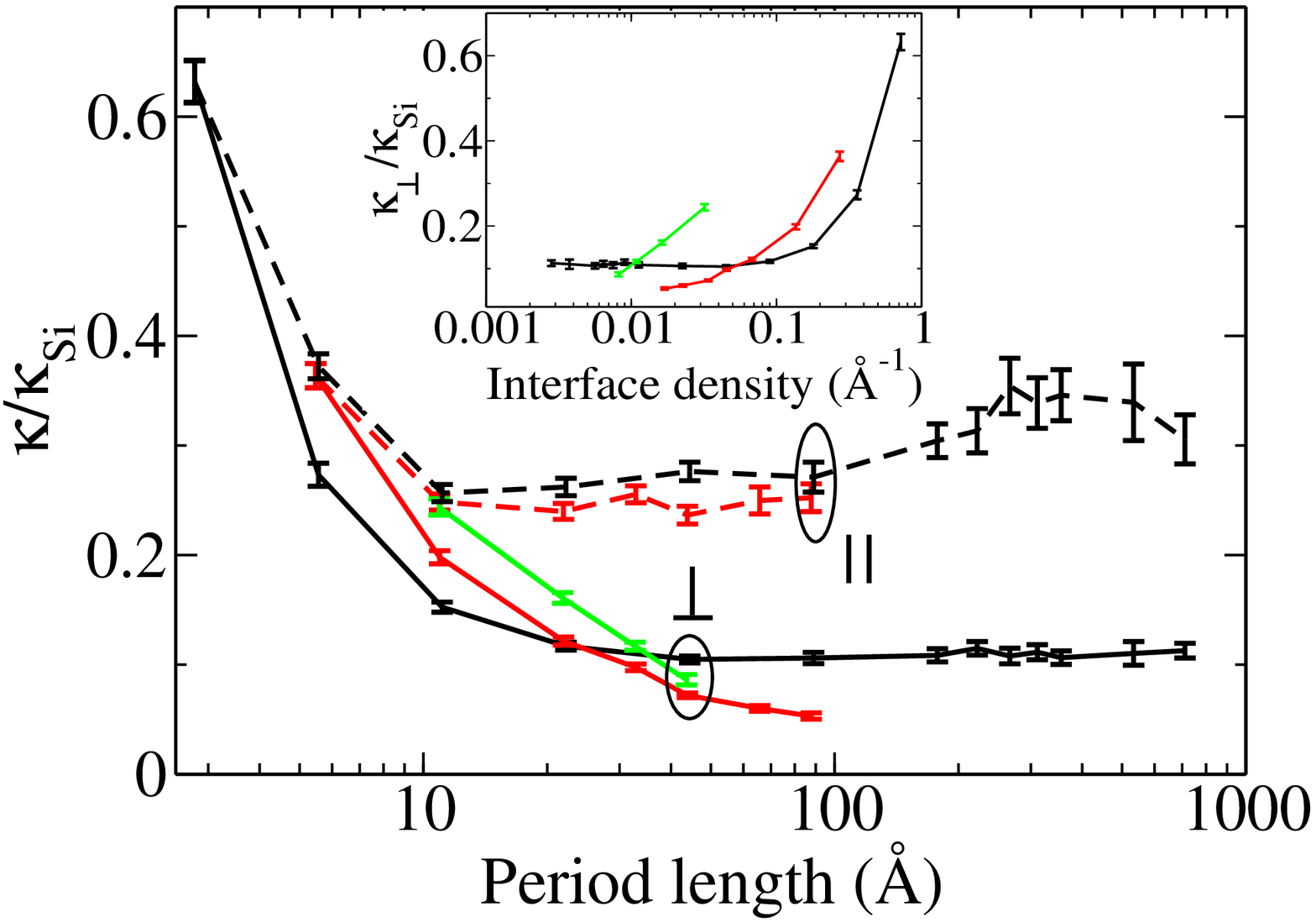}
\caption{The ratio of the thermal conductivity computed using the Monte Carlo method in the cross-plane (solid line) and in-plane direction (dashed line) and the thermal conductivity of bulk Si at room temperature for Si-Ge planar (SL, black line), nanowire (NW SL, red line) and nanodot superlattices (ND SL, green line) as a function of the superlattice period. Inset: the same ratio in the cross-plane direction as a function of the interface density (interfacial area per unit volume).}
\label{fig3}
\end{figure}

Our atomistic calculations of phonon lifetimes permit a detailed analysis and understanding of the observed thermal conductivity dependence on the period length for all types of SLs. The mass mismatch between SL layers leads to a flattening of the system's phonon dispersions with respect to those of the bulk, and to the formation of minibands~\cite{prl84-927}. Our results for planar SLs show that this change in the phonon dispersion is accompanied by a reduction of the group velocities $v_g$ as $L$ increases (also reported in Refs.~\cite{prl84-927,prb56-10754,prb60-2627,prb70-081310,prl101-105502}), and finally their saturation for $L>90$ \AA. 
Phonon lifetimes $\tau$ exhibit a minimum for $L\approx 11$ \AA, and the two effects combined lead to a minimum in the $\kappa$. We found similar trends for $v_g$ and $\tau$ in the case of NW SLs when heat propagates in the in-plane direction. On the other hand, the cross-plane group velocities and $\kappa$ of ND and NW SLs decrease as $L$ increases and
do not saturate for the period lengths 
$<90$ \AA, in contrast to planar SLs. Our calculations show that the rate of the $v_g$ reduction decreases as the size of Ge wires and dots increases, which indicates that the saturation will occur for the sizes larger than those considered here.

We now turn to the discussion of the effect of 
dimensionality on the computed thermal conductivity. As shown in Fig.~3, planar SLs have the lowest $\kappa_{\perp}$ for a fixed short period.
For periods longer than $4$ nm, a cross-over occurs and $\kappa_{\perp}$ of planar SLs exceeds that of NW and ND SLs with the same $L$ (by up to a factor of $2$
for NWs with $L\approx 9$ nm). This result suggests that ND and NW SLs with sufficiently large periods are 
better candidates for thermoelectric materials than planar SLs of similar dimensions. The observed effect of dimensionality on $\kappa_{\perp}(L)$ could not be observed in previous atomistic studies, due to the computational difficulties of standard atomistic BTE methods to treat NW and ND SLs. 
\begin{figure}
\includegraphics[width=0.7\textwidth]{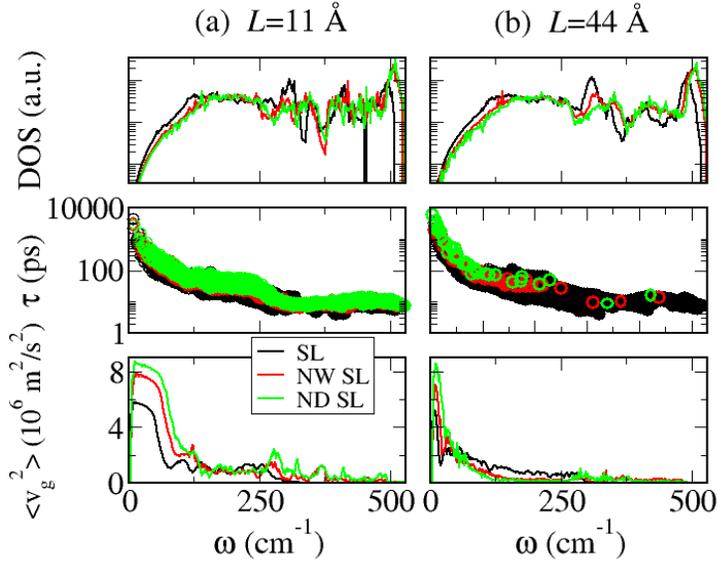}
\caption{Upper, middle and lower panels show the density of states, lifetimes and cross-plane averaged squared group velocity, respectively, as a function of the phonon frequency for planar (SL, black line), nanowire (NW SL, red line) and nanodot (ND SL, green line) superlattices at 300 K. Left (right) panel shows the results for the period $L$ of $11$ \AA~($44$ \AA). Lifetimes for nanowire and nanodot superlattices with $L=44$ \AA~were obtained using Monte Carlo integration, while all other results were computed using exact integration.}
\label{fig5}
\end{figure}

Our calculations show that for the SLs with sharp interfaces considered in our study, the observed cross-over in $\kappa_{\perp}(L)$ for  $L<10$ nm is not simply determined by a change in the interface density, which was proposed to have the dominant effect on the $\kappa$ of SLs with diffuse interfaces and $L>10$ nm in previous model calculations~\cite{advmat19-1043,prb69-195316,jht130-042410,jjap50-035201}. The inset of Fig. 3 shows that SLs of different dimensionality with the same interface density do not have the same $\kappa$. The cross-over is instead the consequence of the changes  of group velocities, lifetimes and density of states, as shown in Fig. 4. Fig. 4(a) (Fig. 4(b)) shows our results for SLs with the period length of $11$ \AA~($44$ \AA), corresponding to the case of lowest (largest) $\kappa_{\perp}$ found for planar SLs.
We note that planar SLs have $50$\% of Ge content, while NW and ND SLs have $\approx 16$\% and $\approx 6$\%, respectively. 
The SLs with the larger Ge concentration have larger density of states~\cite{footnote_si} 
at low $\omega$ and lower lifetimes, as shown in Fig. 4(a).
For short $L$, the group velocities~\cite{footnote_si} 
of planar SLs are smaller than those of NW and ND SLs.
The lower group velocities and lifetimes of planar SLs lead to a lower $\kappa_{\perp}$ than that of NW and ND SLs for short $L$, in spite of the larger density of states in planar SLs. However, as $L$ increases, the group velocities of NW and ND SLs substantially decrease for almost all frequencies, while those of planar SLs slowly saturate (see the lower panel of Fig. 4(b)). In contrast, the density of states and lifetimes exhibit much less pronounced changes with increasing $L$. 
Unlike group velocities, their values at any particular $\omega$ depend on a range of other frequencies (Eq. (1); also Eq. (3)~\cite{footnote_si}), and they are thus less sensitive to the phonon band structure details and dimensionality.
For sufficiently large $L$, the reduced group velocities and the low density of states of NW and ND SLs lead to a larger reduction in $\kappa_{\perp}$ than that of planar SLs, in spite of the larger NW and ND lifetimes. Hence the $\kappa_{\perp}$ of NW and ND SLs becomes lower than for planar SLs. Our atomistic analysis thus demonstrates that, to understand the effect of dimensionality on the SL thermal conductivity, it is necessary to go beyond model calculations using frequency averaged bulk phonon dispersions and lifetimes, and fully capture their atomistic details.
 
Our results for small period SLs show that in the temperature range $200\--600$ K where the BTE formalism is applicable and three-phonon scattering processes dominate, the $\kappa$ is roughly inversely proportional to the temperature, similarly to bulk Si and Ge. Therefore, our conclusions for $300$ K should be valid in the whole temperature range.

We note that our calculated $\kappa/\kappa_{\text{Si}}$ ratios 
for planar SLs
are $3\--4$ times larger than the experimental values~\cite{apl70-2957,slsmicro28-199}. Similar ratios have been reported in a recent BTE study using first principles Hamiltonians for Si and Ge~\cite{nl12-5135}. This indicates that interface roughness and the presence of defects may play a prominent role in determining the measured values of $\kappa$.

In summary, we studied the effect of dimensionality on the thermal conductivity of planar superlattices with Si and Ge layers of equal thickness, and arrays of Ge nanowires and nanodots embedded in Si matrix with similar sizes, grown in the [001] direction. 
We predicted a cross-over in the cross-plane thermal conductivity ($\kappa_{\perp}$) behavior as a function of dimensionality for temperatures $\sim 300$ K: planar superlattices conduct heat less (more) efficiently than nanowires and nanodots for periods shorter (longer) than several nm. 
Our calculations on planar superlattices and Ge wires in Si with different Si/Ge layer thickness ratios~\cite{footnote_si} indicate that the $\kappa_{\perp}$ dependence on dimensionality in this class of materials remains qualitatively the same as reported here. The period length where the cross-over occurs decreases as the thickness ratio decreases, and reaches the minimal period length for the ratios $\sim 1:3$. On the other hand, unlike Ge wires in Si, Si wires embedded in Ge always have a lower $\kappa_{\perp}$ than their planar counterparts with equally thick Si and Ge layers due to a higher Ge concentration~\cite{footnote_si}.
Our findings may be useful to establish design rules for nanostructures with desired thermal transport properties e.g. for applications requiring low thermal conductivity (thermoelectric conversion processes).
The results presented here were obtained using a 
fully atomistic Monte Carlo method to solve the Boltzmann transport equation in the relaxation time approximation, with greatly enhanced efficiency and accuracy. 


We thank Rama Venkatasubramanian and \'Eamonn Murray for useful discussions.
This research was supported by DOE/BES grant DE-FG02-06ER46462 and computational resources on the Mako cluster of the University of California Shared Computer Center.


\end{document}